\documentstyle[emulateapj,apjfonts]{article}

\def\kms{km\, s$^{-1}$}
\newcommand{\Msun}{$M_{\odot}$}

\def\h65{$h_{65}$}
\def\fesc{$f_{esc}$}

\def\lya{Ly$\alpha$}
\def\h2{H$_{2}$}  
\def\fevol{$f_{\rm evol}$}  

\newcommand\fq[1]{$f_{\rm Q_{#1}}$} 
\input epsf

\begin{document}

\title{PROBING THE FIRST STARS WITH HYDROGEN AND HELIUM RECOMBINATION
EMISSION}
\author{JASON TUMLINSON, MARK L. GIROUX, and J. MICHAEL SHULL\altaffilmark{1}}
\affil{Center for Astrophysics and Space Astronomy, \\ 
Department of Astrophysical and Planetary Sciences, \\ 
University of Colorado, CB 389, Boulder, CO, 80309 \\ Electronic Mail:
(tumlinso, giroux, mshull)@casa.colorado.edu} \altaffiltext{1}{Also at JILA,
University of Colorado and National Institute of Standards
      and Technology.} 

\begin{abstract}

Unusual patterns of recombination emission from gas ionized by
metal-free stars may distinguish early star-forming galaxies from their
present-day counterparts.  This pattern arises from the harder
ionizing spectrum expected from metal-free stars, which strongly
enhances the strength of \ion{He}{2} recombination lines. Our
calculations indicate that line fluxes of \ion{He}{2} $\lambda$1640 and
$\lambda$4686 are sufficiently large to be detected by narrowband and
spectroscopic searches for high-redshift emission-line sources at $z
\sim 5$ using current instruments. An unknown fraction of Ly$\alpha$
emitters may harbor low-metallicity or metal-free stars.  As the
predicted \ion{He}{2} $\lambda$1640 flux is comparable to and may
exceed hydrogen Ly$\alpha$, searches for high-redshift galaxies should
consider \ion{He}{2} recombination lines as possible identifications
for single emission lines in observed spectra.  Spectra of metal-free
stars may show both \ion{H}{1} and \ion{He}{2} emission lines,
improving the constraints on their redshift and identification.  We
assess the considerable uncertainties that affect our expectations for
the detection and identification of true first-generation stars with
present search techniques, including the role of stellar mass loss in
spectral evolution and the confusion of ionization by primordial
stellar sources and AGN in the early universe.  \end{abstract}


\section{INTRODUCTION}

The first stars have long attracted interest from a cosmological point
of view, and recent detailed studies of their properties have renewed
interest in their cosmological importance.  Within the constraints of
big-bang nucleosynthesis, these stars must have been metal free, which
modifies their structure in fundamental ways:  metal free stars have
smaller radii, hotter cores, and higher effective temperatures than
their counterparts of equal mass but finite metallicity (Tumlinson \&
Shull 2000). These stars are leading candidates for the sources that
reionized the intergalactic medium (Gnedin 2000; Haiman \& Loeb 1997),
and they produced the first heavy elements that enriched subsequent
stellar generations.  Tumlinson \& Shull (2000) found that these stars,
here called Population III or simply metal-free, have harder spectra
and emit more ionizing photons in the \ion{He}{2} ionizing continuum
than do stars of typical solar or sub-solar metallicity.  Thus, they
may dominate \ion{He}{2} reionization in the high-redshift universe.

Distinctive spectral features of their surrounding nebulae may provide
a means of detecting the first stars, during and after the epoch of
reionization.  Tumlinson \& Shull (2000) noted that the hard spectra of
Pop III stars produce large \ion{He}{3} regions, which may emit
detectable \ion{He}{2} recombination emission.  Recombination lines of
\ion{He}{2} at $\lambda$1640 ($n = 3 \rightarrow$ 2), $\lambda$3203 ($n
= 5 \rightarrow$ 3), and $\lambda$4686 ($n = 4 \rightarrow$ 3) are
particularly attractive for this purpose, because they suffer minimal
effects of scattering by gas and decreasing attenuation by intervening
dust.  Thus, an assessment of the near-term prospects for discovering
metal-free stellar sources is timely.  However, uncertainties about the
primeval initial mass function (IMF), the importance of stellar mass
loss at zero metallicity, and the features of stellar evolution must be
addressed.

In this {\em Letter}, we consider the possibility of detecting
metal-free stars using \ion{He}{2} recombination emission, and the
uncertainties inherent in this technique.  Main-sequence mass loss, not
well understood from an evolutionary standpoint, can affect the
detectability of the first stars via nebular emission lines.  Emission
line ratios and line profiles may discriminate between stars and active
galactic nuclei (AGN) as sources of ionization.  We address these
issues in the context of existing tracks and assess their importance to
the detectability of metal-free stars.

In \S~2 we derive an expression for the observed \ion{He}{2}
emission-line flux of a metal-free stellar population in terms of the
star formation rate and evolutionary properties. In \S~3 we use
evolutionary tracks and model atmospheres to assess the effects of
stellar evolution on the detectability of metal-free stellar
populations.  In \S~4 we compare our predictions to current
observational searches for high-redshift stellar populations, including
present-day emission-line techniques, and we discuss the ambiguity
between emission lines produced by metal-free stars and AGN.  In \S~5
we summarize our results and comment on directions for future work.

\section{Emission Lines from Metal-Free Stars} 

We begin by estimating a relationship for the observed \ion{He}{2}
emission-line flux from metal-free stars as a function of star
formation rate (SFR) and stellar evolutionary parameters.  Using the
zero-age structure and atmosphere models presented by Tumlinson \&
Shull (2000), we use recombination theory to estimate the total
luminosity in the \ion{He}{2} emission lines.  We assume that
few \ion{He}{2} ionizing photons escape the galaxies where they are
produced (\fesc\ $\simeq$ 0) and that the sources themselves contain no
dust.  We use the Kennicutt (1983) law to relate the luminosity of the
H$\alpha$ line to the star formation rate, L(H$\alpha$) = $(1.12 \times
10^{41}$ erg s$^{-1})$ (SFR/M$_{\odot}$ yr$^{-1}$).  For case B
recombination at $T$ = 20,000 K (corresponding to a higher nebular
temperature in low-metallicity gas) and $n_{\rm He III}/n_{\rm HII}$ =
0.0789, we find $j_{4686}/j_{\rm H\alpha}$ = 0.33, $j_{1640}/j_{\rm
H\alpha}$ = 2.3, and $j_{3203}/j_{\rm H\alpha}$ = 0.14 (Seaton 1978).
Kennicutt (1983) assumed a Salpeter-like IMF and used stellar evolution
tracks for Population I massive stars. We assume a similar IMF and
include a factor, \fevol\ (of order unity), designed to account for the
time evolution of the stellar ionizing continuum radiation, and defined
to be unity if the evolution in \ion{He}{2} ionizing photon rate for
metal-free stars is identical to the evolution of \ion{H}{1} ionizing
photons used by Kennicutt (1983).  The factor \fevol\ can be evaluated
using Pop III evolutionary tracks coupled with model atmospheres (see
\S~3).  Using the results of Tumlinson \& Shull (2000), we scale
the \ion{He}{2} ionizing photon production of zero-metallicity stars to
the \ion{H}{1} ionizing photon production implicit in the Kennicutt
(1983) relation.  We find that Pop III stars produce $10^{-1.1}$ as
many \ion{He}{2} ionizing photons as the Pop I stars produce \ion{H}{1}
ionizing photons.  Scaling the luminosity of \ion{He}{2} $\lambda$1640
to H$\alpha$, we find:
\begin{equation}
L_{1640} = (4.2\times 10^{40} \; {\rm erg\; s^{-1})\; }f_{\rm evol} \left(\frac{\rm SFR}{M_{\odot}
\rm yr^{-1}}\right).   
\end{equation}
\noindent For deceleration parameter $q_{0}$ = 0.5, the flux of this 
source is: 
\begin{equation} 
F_{1640}(z) = \frac{(4.1 \times 10^{-18} \; {\rm erg\; cm^{-2}\; s^{-1}}) h_{65}^2 
f_{\rm evol} \rm SFR}{[(1+z) -(1+z)^{\frac{1}{2}}]^2} , 
\end{equation} 
\noindent where $h_{65}$ is the Hubble constant in units 65 \kms\
Mpc$^{-1}$.  The flux of \ion{He}{2} $\lambda$4686 is 7.1 times lower
than $F_{1640}$, and the flux of $\lambda$3203 is 16.2 times lower.
(Hereafter we concentrate on the two stronger lines.) This relation is
plotted in Figure 1 for SFR = 5 and 20 \Msun\ yr$^{-1}$ and \fevol\ =
0.4 and 2.0; these choices of \fevol\ are justified in \S~3.  We
also plot detection limits of three recent emission-line searches for
high-redshift galaxies. Figure 1 shows that, even for conservative
estimates of \fevol, metal-free stellar populations may be detectable
to $z = 2 - 5$ for a reasonable range of SFR. If a detection can be
made and its redshift accurately measured, the star formation rate can
be constrained if \fevol\ is carefully calibrated by stellar evolution
models. 

\begin{figure*} 
\centerline{\epsfxsize=0.7\hsize{\epsfbox{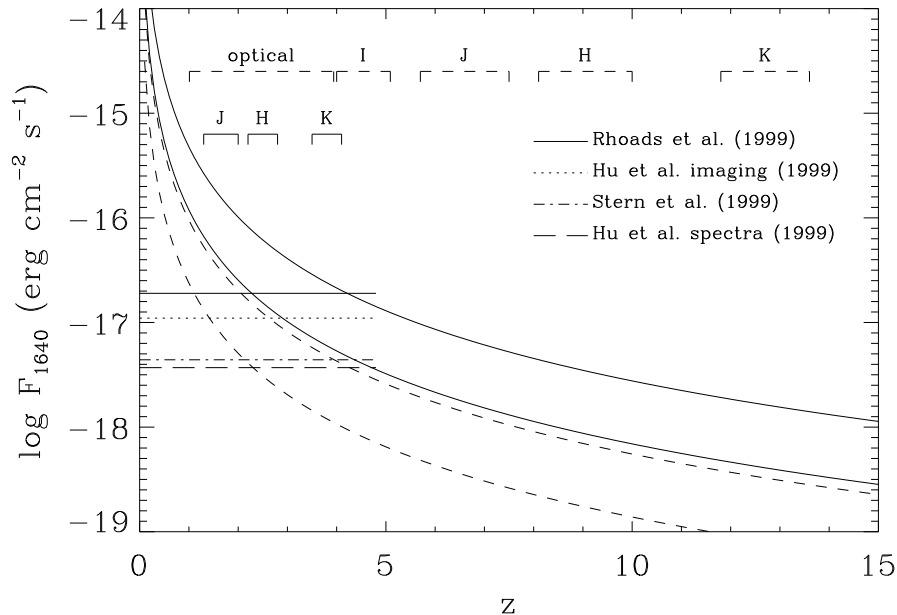}}}
\vspace{-0.2in} 
\figcaption{Flux of He~II $\lambda$1640 line as a function of source
redshift (eq.~2). The dashed and solid curves correspond to \fevol\ =
0.4 and 2.0, respectively. The upper and lower curves in each pair
correspond to star formation rates 20 and 5 \Msun\ yr$^{-1}$.  The flux
for $\lambda$4686 is 7.1 times lower than $F_{1640}$. The horizontal
solid, dotted, dash-dot, and dashed lines show the limits of current
Ly$\alpha$ emission-line surveys, as described in the text. The flux
limits have been corrected for the energy difference between 1216
\AA\ and 1640 \AA.  At the top, we mark the ranges of redshift probed
by the He~II $\lambda$1640 (above) and $\lambda$4686 (below) lines for
optical and infrared searches. Optical searches for $z = 1 - 5$ are
feasible now, and infrared searches may be possible in the near future.
\label{fig1}}
\medskip 
\end{figure*}

\section{Stellar Evolution Effects} 

Stellar evolution at zero metallicity is uncertain and may enhance or
diminish the detectability of metal-free stars.  Tumlinson \& Shull
(2000) estimated the ionizing photon production from zero-age main
sequence (ZAMS) metal-free stars of mass 2 -- 90 \Msun\ using static
stellar structure models and NLTE model atmospheres.  The \ion{He}{2}
ionization produced by these stars is a direct result of their high
effective temperatures.  However, published evolutionary tracks of
metal-free stars (Castellani et al.  1983; Chieffi \& Tornambe 1984)
show that these stars may evolve to cooler temperatures and larger
radii over their lifetimes if they do not experience mass loss.
Because mass loss may play a significant role in the spectral evolution
of the star, it produces uncertainty in the interpretation of
emission-line diagnostics.

With model atmospheres similar to those presented by Tumlinson \& Shull
(2000), Hubeny, Lanz, \& Heap (2000) argue that line-blanketed,
radiation pressure-driven winds are not initiated for stars with $Z
\lesssim 0.001$, owing to the relative lack of metal line-blanketing in
their atmospheres.  Kudritzki (2000) draws similar conclusions.
However, these groups have not, as yet, extended their models to the range
$T_{\rm eff} \geq 60000$ K occupied by metal-free massive stars.

Because these stars radiate near the Eddington limit, they may
drive winds with electron-scattering opacity.  El Eid et al. (1983)
used an empirical mass-loss prescription from Chiosi (1981) that scaled
$\dot{M}$ to the Eddington luminosity with no explicit dependence on
metallicity.  They found typical mass loss rates of $\dot{M} \sim
10^{-5}$ \Msun\ yr$^{-1}$ for stars of 80 - 500 \Msun, with little
dependence on luminosity.  At these rates, a star with initial mass 80
\Msun\ will fall to 40 \Msun\ by the end of hydrogen burning, when it
begins an excursion to hotter $T_{\rm eff}$ in the HR diagram. As its
outer layers are lost, hotter regions below are exposed, and the
interior structure adjusts to the changing mass and photospheric
conditions.  At late times, $T_{\rm eff}$ increases by 0.2 dex over
their ZAMS, increasing the production of He II ionizing photons 
by an order of magnitude. 


To evaluate the importance of stellar evolution and mass loss on the
production of \ion{He}{2} ionizing photons, we make two limiting
assumptions:  (1) mass loss from Population III stars is negligible,
with no effect on their evolution; (2) mass loss is important and
affects the structural and spectral evolution of the star.  We use the
25 \Msun\ track calculated by Castellani et al.~(1983) to represent the
first case, and the 80 \Msun\ mass-loss track by El Eid et al.~(1983)
for the second.  These tracks are not intended to replace the stellar
models on which our flux estimates are based. Rather, they represent
extremes of mass loss used to constrain the parameter space of \fevol.
If mass loss is unimportant to the evolution of metal-free stars, then
the general trend to lower $T_{\rm eff}$ seen in the 25 \Msun\ track
should hold for stars at all mass and will favor low values of \fevol.
Conversely, if mass loss substantially affects the later evolutionary
phases of metal-free stars, then the trends toward higher $T_{\rm eff}$
at late time will favor larger \fevol.

Model atmospheres were calculated with the TLUSTY code (Hubeny \& Lanz
1995) for a set of points along the evolutionary tracks, placed to
capture the time evolution of $T_{\rm eff}$ and surface gravity
$g$. We define a time-averaged flux parameter, 
\begin{equation}
f_{Q_{i}} = \frac{\int_{t_{\rm PSN}} Q_i(t) dt }{Q_i(0) t_{\rm PSN}} ,
\end{equation}
where $Q_{i = 0,1,2}$ are the ionizing photon production rates for
\ion{H}{1}, He~I, and \ion{He}{2}, respectively, and $t_{\rm PSN}$ is
the pre-supernova lifetime of the star.  The 25 \Msun \ star
evolves to lower $T_{\rm eff}$ during H burning and experiences a burst
of He~II ionizing photons near the onset of He burning. The 80
\Msun\ star loses over half of its mass during its lifetime and makes
two brief excursions to high $T_{\rm eff}$ and high $Q_2$. For the
constant-mass (25 \Msun) track, we derive \fq{0} = 1.3, \fq{1} = 1.1,
and \fq{2} = 0.32.  For the 80 \Msun\ track with mass loss, we find
\fq{0} = 1.2, \fq{1} = 1.1, and \fq{2} = 1.80.  For comparison, we
estimate that \fq{0} = 0.7, \fq{1} = 0.4, and \fq{2} = 0.3 are
characteristic of the Chiosi, Nasi, \& Sreenivasan (1978) Population I
tracks used by Kennicutt (1983).  Because \fevol\ compares the
evolutionary trends in the He II ionizing flux from metal-free stars
with the time average in H I ionizing flux from Pop I, our estimate is
given by \fevol\ = \fq{2} (Pop~III)/\fq{0} (Pop~I).  Thus, on
timescales long compared to the evolution of massive stars, \fevol\ may
range from 0.4 to 2.6. The two limiting cases imply different behavior
on short timescales.  For $\dot{M} = 0$, the \ion{He}{2} emission fades
and H~I emission brightens over time as the stars evolve to cooler
temperatures.  For $\dot{M} \neq 0$ the \ion{He}{2} emission from a Pop
III cluster increases over time, as stars of successively lower mass
exhaust their central hydrogen and make excursions to hotter $T_{\rm
eff}$.

Our calculation illustrates a key feature of the ionizing photon
production by the first stars: emission-line diagnostics evolve with
their stellar populations, and this must be considered in the planning
and interpretation of observational searches.  A single snapshot of a
metal-free stellar population in its \ion{He}{2} recombination lines
would not distinguish between the $\dot{M} = 0$ and $\dot{M} \neq 0$
cases, because of the degeneracy in the parameters \fevol\ and SFR
(eq.~2). Measuring the star formation rate of metal-free stars with
\ion{He}{2} emission requires that \fq{2} be carefully calibrated by
stellar evolution calculations.

The evolution of the ionizing photon production may provide a means of
distinguishing between the two cases presented above.  The time
evolution of $Q_{i}$ for the limiting cases of $\dot{M} = 0$ and
$\dot{M} \neq 0$ illustrates a key difference between the two
possibilities. The 25 \Msun\ star at constant mass exhibits most
efficient \ion{He}{2} ionization at the beginning of its H-burning
main-sequence, giving $Q_2/Q_0$ = 10$^{-4}$. The 80 \Msun\ star with
$\dot{M} \neq 0$, by contrast, achieves the maximum value $Q_2/Q_0$ =
0.40 at the end of its life.  This ratio is larger than the range of
$Q_2/Q_0$ = 0.05 -- 0.12 for the most massive stars studied by
Tumlinson \& Shull (2000) and Bromm, Kudritzki, \& Loeb (2000).  Thus,
the two extreme cases can be distinguished by observed \ion{He}{2} to
\ion{H}{1} ionization ratios that exceed those achieved by
constant-mass stars.

\section{Comparison With Other Emission-Line Diagnostics} 

The relatively bright \ion{He}{2} emission-line fluxes predicted for
metal-free stars raise the possibility of detection with present-day
broadband and spectroscopic searches. In particular, we examine the
possibility that \ion{He}{2} $\lambda$1640 emitters have already been
observed and mistaken for Ly$\alpha$ emitters.  Recent searches for
high-redshift galaxies use narrow-band imaging and spectroscopic
techniques.  These emission lines are believed to be Ly$\alpha$ because
deep spectra reveal no other galactic emission lines ([O~III]
$\lambda$5007, [O~II] $\lambda$3727, H$\alpha$), and these
identifications for the single line can be eliminated by the absence of
corresponding emission.

In Figure 1 we plot the detection limits of three current Ly$\alpha$
searches for high-redshift galaxies.  Hu, Cowie, and McMahon (1999)
quote a 5$\sigma$ detection limit $1.5 \times 10^{-17}$ erg cm$^{-2}$
s$^{-1}$ for narrowband imaging and $5.0 \times 10^{-18}$ erg cm$^{-2}$
s$^{-1}$ (5$\sigma$) for spectroscopic searches.  In a 1.5 hour Keck
exposure, Stern et al.~(2000) achieved a limiting flux $6.0 \times
10^{-18}$ erg cm$^{-2}$ s$^{-1}$ in a 1$''$ slit for an unresolved
line. Rhoads et al. (2000) achieved detections ranging between $1.8 -
2.6 \times 10^{-17}$ erg cm$^{-2}$ s$^{-1}$ using combinations of
imaging and spectroscopy.  As shown in eq. 2, such a detection of
\ion{He}{2} lines would measure the product of SFR and \fevol.
Therefore, careful calibration of \fevol\ with stellar evolution models
could break the degeneracy and independently constrain the star
formation rate.

Hu, McMahon, \& Cowie (1999) reported the detection of a single line in
the spectrum of a luminous galaxy, using a combination of narrowband
imaging and spectroscopic observations.  Identifying this line as
Ly$\alpha$, they infer $z = 5.74$, a total line flux $1.7 \times
10^{-17}$ erg cm$^{-2}$ s$^{-1}$ at 8195 \AA, and a star formation rate
of 19 \Msun\ yr$^{-1}$.  If this line is \ion{He}{2} $\lambda$1640 at
$z = 4.0$, the implied SFR would be 80, 32, and 16 \Msun\ yr$^{-1}$ for
\fevol ($Q_2$) = 0.4, 1.0, and 2.0, respectively.  Typical Lyman-break
galaxies have inferred SFR $\simeq 10 - 100$ \Msun\ yr$^{-1}$ (Steidel
et al. 1999). Thus, the SFR by itself does not rule out identification
as \ion{He}{2} $\lambda$1640 for implied values of \fevol\ consistent
with the $\dot{M} \neq 0$ case.  The implied total flux in
$\lambda$4686 is $2.3 \times 10^{-18}$ erg cm$^{-2}$ s$^{-1}$ at 2.34
$\mu$m, below the K-band sensitivity limits of the Hu et al.  (1999)
survey.  The flux predictions in \S~2 suggest that this line might be
\ion{He}{2} $\lambda$1640 at $z = 4.0$, were it not for the presence of
a spectral break across the line produced by intergalactic
\lya\ absorption.  The absence of the break may provide key
spectroscopic evidence for identifications of \ion{He}{2} recombination
lines.  Without a break, the U, B, and V dropout techniques commonly
used to discover high-redshift galaxies (Steidel et al.~1999) may not
select these objects for spectroscopic followup.  Because $Q_{0}$ for
Pop III is similar to that of Pop I (Tumlinson \& Shull 2000), the
observed fluxes of He II $\lambda$1640 and \ion{H}{1} Ly$\alpha$ should
be comparable. The predicted line fluxes and the separation of
Ly$\alpha$ and \ion{He}{2} $\lambda$1640 by 424(1+$z$) \AA\ suggests
that some detections may show two emission lines, confirming
identification and redshift in the range $z = 2 - 5$.  Future
spaceborne missions such as {\em NGST} may also be used to detect
high-redshift AGN and metal-free stars via associated \ion{He}{2}
$\lambda$1640 and $\lambda$4686 nebular emission (Oh, Haiman, \&
Rees~2001).

It remains an issue, however, how the detection of a single nebular
line would permit one to discriminate between stellar and AGN sources
of ionizing radiation.  Intrinsic line widths may not be sufficient to
resolve the confusion. Recent surveys of Wolf-Rayet galaxies (Schaerer
et al.~1999 and references therein) show broad \ion{He}{2} $\lambda$4686
emission from stellar winds and narrow nebular emission. These lines
are usually detected at fluxes of 1 -- 4\% of the observed H$\beta$
fluxes from these galaxies, and the broad wind features are generally
brighter than the nebular emission.  Spectroscopic Ly$\alpha$ surveys
(Hu et al.~1999) typically use low resolution, R $\sim$ 1000, to
achieve their required sensitivity.  Identifying stellar and AGN
ionization may require resolution of $\sim$100 km s$^{-1}$ to
distinguish broad, cuspy line profiles of AGN lines from narrow nebular
emission due to massive stars.  The \ion{He}{2} to \ion{H}{1} line ratios for
metal-free stars may provide a further discriminant. The $\dot{M} \neq 0$
case (\S~3) gives a maximum value of $Q_{2} / Q_{0}  = 0.40$. For a
power-law AGN spectrum, $f \sim v^{-\alpha}$, this ratio is $Q_{2} /
Q_{0} = 4^{-\alpha}$. For the composite radio-quiet AGN spectrum
($\alpha = 1.8$) from Zheng et al.~(1997), this ratio is $Q_{2} /
Q_{0} = 0.08$.  However, distinguishing metal-free stars from AGN with
harder spectra than the power-law composite may be more difficult. If
mass loss is important to the evolution of metal-free stars, this
maximum ratio technique may provide a convenient means of
distinguishing them from AGN in the early universe. The prospects for
distinguishing stellar and AGN ionization improve for cases where lines
of both \ion{H}{1} and \ion{He}{2} are observed.

As a potential discriminant between AGN and metal-free stars, IGM
ionization ratios of \ion{He}{2} to \ion{H}{1} may show different
signatures of AGN or stellar ionization.  However, the hard AGN-like
ionizing continuum of metal-free stars may confuse the issue.  Fardal,
Giroux, \& Shull (1998) define the column density ratio $\eta$ =
N$_{\rm HeII}$/N$_{\rm HI}$.  Ultraviolet studies of the \ion{He}{2}
Gunn-Peterson absorption toward quasars at $z = 2.7 - 3.3$ (Davidsen,
Kriss, \& Zheng 1996; Reimers et al. 1997; Hogan, Anderson, \& Rugers
1997) suggest that the IGM reionizes in \ion{He}{2} at $z \approx 3$.
The ionizing spectrum required to match the optical depths, $\tau_{\rm
HI}$ and $\tau_{\rm HeII}$, at $z \leq 3$ is relatively soft, with
$\eta = 100 - 200$, in contrast to the harder spectra from unfiltered
AGN and metal-free stars, which yield $\eta \approx 30$.  If metal-free
stars are implicated in the \ion{He}{2} reionization, then
observational searches for the characteristic \ion{He}{2} emission
lines ($\lambda$1640 and $\lambda$4686) from starburst galaxies at $z =
3 - 5$ would be extremely helpful.  These lines should be measured
together with high-ionization metal absorption lines (\ion{C}{4},
\ion{S}{4}), to assess any shift in the quality of the ionizing
spectrum at $z \approx 3$ (Songaila \& Cowie 1996).  Discriminating
stellar from AGN sources may depend on detailed line ratio predictions
from accretion models (Hubeny et al. 2000) and stellar population
synthesis. The maximum line ratio technique sketched above must be
extended to account for IMF and evolution effects, but it holds promise
if complete \ion{H}{1} and \ion{He}{2} recombination spectra are
observed.

\section{Summary} 

In summary, our main conclusions are the following: 

\begin{itemize} 
\item[1.] Predicted fluxes for \ion{He}{2} emission lines from nebulae
ionized by metal-free stars are comparable to the detection limits of
present-day narrowband and spectroscopic searches for high-redshift
galaxies. Observational searches for these stars at $z \sim 3 - 5$ are
feasible now at optical wavelengths.  Higher redshift surveys may need
to await space-based infrared instruments.

\item[2.] Mass loss by metal-free stars may restrict or enhance their
detectability with emission-line techniques.  Careful calibration with
detailed evolutionary tracks and theoretical work on mass loss from
metal-free stars are necessary to refine this program.

\item[3.] If multiple line detections are made for a single object,
\ion{He}{2} and \ion{H}{1} emission lines may provide important
diagnostics of the stellar parameters. These line ratios may provide a
means of distinguishing stellar and AGN sources even if the line
profiles are unresolved.

\end{itemize} 

\acknowledgements 

We are grateful to Peter Conti for suggesting \ion{He}{2} $\lambda$1640
as a signature of hot star ionization, and to John Stocke for helpful
comments about the manuscript.  This work was supported in part by
astrophysical theory grants to the University of Colorado from NASA
(NAG5-7262) and NSF (AST96-17073).

\pagebreak

\end{document}